\documentclass[aps,twocolumn,english,superscriptaddress,citeautoscript,
preprintnumbers,amsmath,amssymb,floatfix,footinbib]{revtex4-1}
\usepackage{hyperref}
\hypersetup{colorlinks=true,linkcolor=red,citecolor=blue}
\usepackage{amsmath}
\usepackage{graphicx}
\usepackage{dcolumn}
\usepackage{float}
\usepackage[utf8]{inputenc}

\usepackage{epstopdf}
\usepackage{siunitx}

\usepackage{fancyhdr}
\begin{document}

\title{Ferromagnetic resonance studies of strain tuned Bi:YIG films}
\author{Ravinder Kumar}
\author{B. Samantaray}
\author{Z. Hossain$^{\star}$}
\affiliation{Condensed Matter - Low Dimensional Systems Laboratory, Department of Physics, Indian Institute of Technology (IIT) Kanpur – 208016, India}
\email{zakir@iitk.ac.in}

\date{\today}
\begin{abstract}
Bismuth-doped Yttrium iron garnet (Bi$:$YIG) thin films known for large magneto-optical activity with low losses still need to get probed for its magnetization dynamics. We demonstrate a controlled tuning of magnetocrystalline anisotropy in Bi-doped Y$_{3}$Fe$_{5}$O$_{12}$ (Bi$:$YIG) films of high crystalline quality using growth induced epitaxial strain on [111]-oriented Gd$_{3}$Ga$_{5}$O$_{12}$ (GGG) substrate. We optimize a growth protocol to get thick highly-strained epitaxial films showing large magneto-crystalline anisotropy, compare to thin films prepared using a different protocol. Ferromagnetic resonance measurements establish a linear dependence of the out-of-plane uniaxial anisotropy on the strain induced rhombohedral distortion of Bi$:$YIG lattice. Interestingly, the enhancement in the magnetoelastic constant due to an optimum substitution of \textit{Bi}$^{3+}$ ions with strong spin orbit coupling does not strongly affect the precessional damping ($\sim2\times10^{-3}$). Large magneto-optical activity, reasonably low damping, large magnetocrystalline anisotropy and large magnetoelastic coupling in Bi:YIG are the properties that may help Bi:YIG emerge as a possible material for photo-magnonics and other spintronics applications.

\vskip 1cm
\end{abstract}

\maketitle
\newpage

Magneto-crystalline anisotropy and Gilbert damping are the crucial parameters for a material to be used in various spin-based device applications\cite{Chang2017,Wang2014,Du,Wang2017}. The emerging field of spintronics promises dense and fast memory architectures, enabling huge data storage and fast information processing\cite{Parkin2004,Fukami,Joshi,Klingler2014,Klingler2015,Grundler,Ganzhorn,Chumak2015,
Makarov,Egel}. The spin current based devices would be highly efficient with almost no thermal losses unlike charge-based electronics and could be used in energy harvesting by recycle of heat waste via spin-caloritronics\cite{Chang2017,Uchida2008,Uchida2010,Bauer,Kirihara,Heremans}. The miniaturization of such concept-device prototypes requires material media in a thin film form, where the magnetic properties can vary significantly due to different film thicknesses, growth induced strains, crystallographic orientation and substrate-film interface reactions. It is essential to have a physical parameter to tune the magnetic anisotropy in thin films while maintaining the precessional damping as-low-as possible. The strain produced in thin films due to substrate-film lattice mismatch serves as a tuning parameter for magnetic anisotropy and can be varied by changing the film thickness. The uniaxial magnetic anisotropy is the main contributing term in a thin film’s total magnetic anisotropy and as the anisotropy field in a ferromagnetic system has one-to-one mapping with the effective magnetization, we tried to establish a relationship between magnetic damping and the strength of effective magnetization for different ferromagnetic systems.
\begin{figure}[t]
\begin{center}
\includegraphics[trim =0mm 5mm 0mm 5mm, width=80mm]{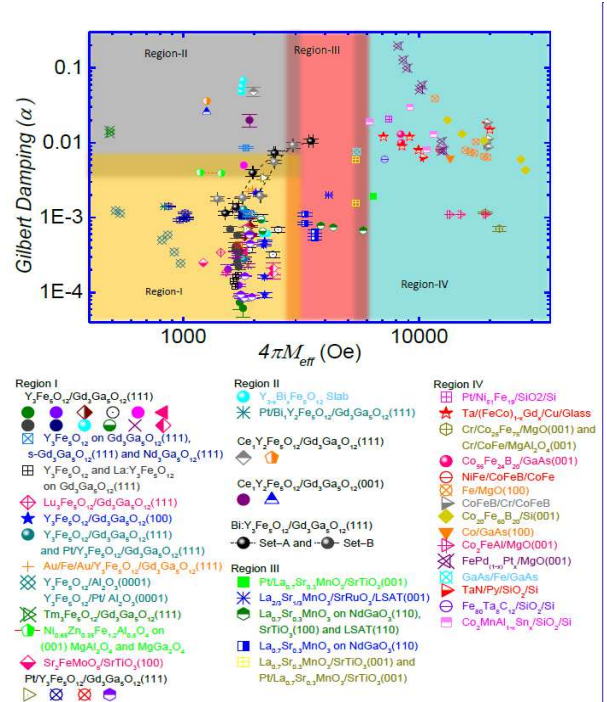}
\setlength{\belowcaptionskip}{-5mm}
\caption{(Color online) Relationship between effective magnetization and Gilbert damping coefficient. Here, we compare some of the interesting work from existing literature; Region I and II: Ferromagnetic insulators in the form of bulk, thin films (polycrystalline and epitaxial); Region III: Conducting-oxides and; Region IV:  Pure metals and metal-alloys. Different regions of interest have been shaded with different colors. Note: References are provided at the end.}\label{F1}
\end{center}
\end{figure}

	In Fig. \ref{F1}, we compile results from existing literature on Gilbert damping ($\alpha$) and effective magnetization $(4\pi M_{eff})$ of different ferromagnetic systems irrespective of the growth (different growth conditions and methods), physical form (thin films or bulk), thickness (in thin films), crystallinity (amorphous or polycrystalline or epitaxial), dopants and other factors (references provided at the end of this paper). Region I is the most exploited one because pure-YIG possesses very low-damping ($\sim10^{-4}$)\cite{Chang2017,Onbasli2014,Chang2014,Howe,Hauser,Krys2017APL}. The application of spin-orbit torque in heavy metals (HM)\cite{Wang2013,Castel,Yang2018,Emori2018,Thiery} and topological insulators (TI)\cite{Lang,Wang2016,Tang2017} capped ferrimagnetic garnet heterostructures show potential to improve the efficiency of magnetic manipulations as it will not shunt a charge current applied to the capped conducting layer\cite{Jermain}. Being an insulating material, only electron’s spin degrees of freedom is allowed, resulting in pure spin current, which is not the case with conducting-oxides (Region III), metals and metal-alloys (Region IV). Besides having the ability to generate pure spin current, the magneto-optical properties of YIG enhances in proportion to Bismuth (Bi) concentration at Yttrium site\cite{Kidoh,Ming-Yau1997,Ming-Yau1999,Helseth}. Due to enhanced magneto-optical activity in the UV, visible and IR regions along with low propagation loss, Bi$:$YIG is a potential candidate in microwave and optical applications such as miniaturization of magnetic field sensors\cite{Kamada,Higuchi,Ch2002,Soibel,Sinha,Mandal2012} and reciprocal transmission devices like isolators and circulators, respectively\cite{Stadler,Onbasli2016,Toshihiro}. It has been well established that the Bi$:$YIG films with in-plane magnetization can serve as basic sensors for magneto-optical imaging of domain formation in magnetic materials, magnetic flux in superconductors, currents in microelectronic circuits and recorded patterns in magnetic storage media\cite{Doros,Koblischka,Zvezdin,Schafer,Vlasko,Egorov}. It is suggestive that the growth parameters optimization is crucial to obtain films with in-plane magnetization and free from effective domain activity\cite{Helseth}\cite{Doros}. Ferrimagnetic insulators with in-plane easy magnetization can also be used to realize spin superfluidity\cite{Takei,Sonin,Upadhyaya,Kim2017,Bunkov}. The coherent condensation of magnons in spin superfluidity offers a unique opportunity to realize long distance coherent superfluid like transport of the spin current, unlike the transport carried by the incoherent thermal magnons which decays exponentially\cite{Upadhyaya}. Recently, coupling of light and spin wave has been demonstrated by irradiating a ferrimagnetic insulator using spatially modulated light beam\cite{Grundler}\cite{Vogel}. This coupling gives rise to a magnonic crystal that shows the capability to be efficiently reprogrammed on demand via heat. The coupling of electromagnetic waves to wave-like excitations in solids (magnons) could also be helpful to reduce all the lateral dimensions by orders of magnitude for on-chip microwave electronics with optically reconfigurable and multifunctional characteristics. Doping pure YIG with Bi improves its sensitivity towards light and makes it pursuable for magneto-optical based device applications. Being a novel material for possible photon-based device applications, it is essential to optimize and investigate the static and dynamic magnetization aspects of this light sensitive material medium (Region II). Bi$:$YIG films with overwhelmingly large magneto-photonic activity coupled with improved magnetic properties will provide a material platform for newly emerging photo-magnonics field.\\
	The importance of Bismuth substituted YIG as a potential material for light based magnonics applications, motivated the studies reported here. In this study, we grow high quality epitaxial Bi:YIG films on GGG(111) crystals using two different growth protocols which allow us to achieve different strain-states induced by rhombohedral distortion due to film-substrate lattice mismatch. We prepared two sets of samples, Set-A and Set-B. Set-A consists of thin Bi:YIG films with large magnetocrystalline anisotropy due to the large magnitude of strain, and, Set-B consists of thick Bi:YIG films with reasonably large strain. Despite being thick, the films from Set-B show large magnitude of strain that leads to large value of magnetocrystalline anisotropy, for an example; the magnitude of uniaxial magnetocrystalline anisotropy field for a 100 \si{\nano\meter} thick film from set-B is larger than a 37 \si{\nano\meter} thin film from Set-A. The Gilbert damping coefficient increases slightly due to strong spin-orbit coupling and inhomogeneity produced by Bismuth doping ($\sim2\times10^{-3}$), but still orders of magnitude smaller compare to metallic films\cite{Okada2014,Song,Guo2014} and are suitable for magnonics\cite{Klingler2014,Klingler2015,Grundler,Ganzhorn,Chumak2015} spintronics\cite{Fukami}\cite{Joshi}\cite{Wolf2001}\cite{Wolf2006} and caloritronics\cite{Chang2017}\cite{Uchida2008,Uchida2010,Bauer,Kirihara} applications. The magnetoelastic constant of Bi:YIG films comes out to be larger than YIG films\cite{Wang2014} due to \textit{Bi}$^{3+}$ substitution which enhances the spin-orbit coupling and hence the magnetoelastic coupling.\\	
\begin{figure*}[t]
\begin{center}
\includegraphics[trim = 0mm 5mm 0mm 3mm, width=172mm]{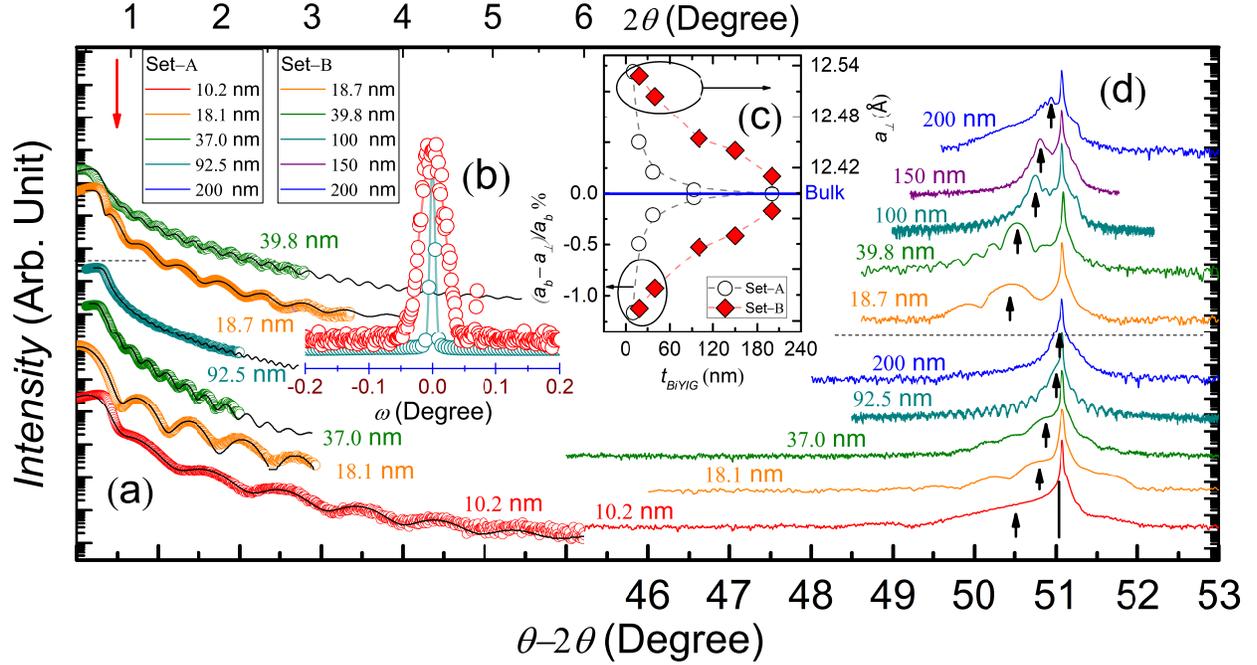}\
\setlength{\belowcaptionskip}{-5mm}
\caption{(Color online) X-ray measurements on Bi:YIG films grown by two different protocols; Panel (a): X-ray reflectivity measurements with fitted data to calculate the thicknesses of different Bi:YIG films. Panel (b): Intensity normalized $\omega$ (Omega) scan profiles with low values of FWHM defines good crystallinity. Panel (c): Thickness dependence of $a_{\bot}$ and percentage strain ($[a_{b}-a_{\bot}]/a_{b}\%$) in the Bi:YIG films from both the sets. Panel (d): X-ray Diffractograms of Bi:YIG films with trails of Laue oscillations suggest high epitaxy.}\label{F2}
\end{center}
\end{figure*}

\begin{figure*}[t]
\begin{center}
\includegraphics[trim = 0mm 5mm 0mm 3mm, width=172mm]{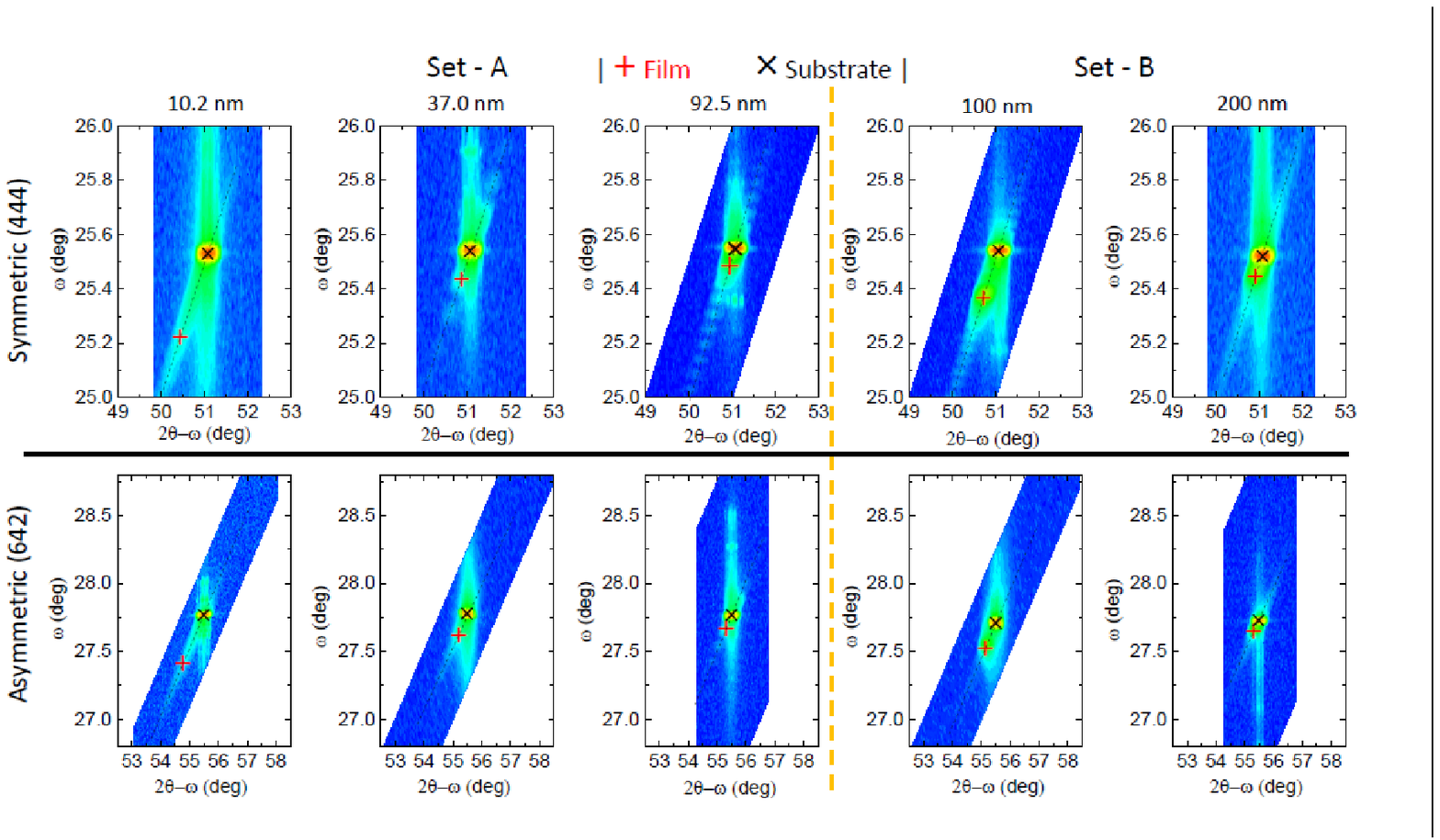}\
\setlength{\belowcaptionskip}{-5mm}
\caption{(Color online) The $\omega$ vs. $2\theta-\omega$, 2-axis maps in $(444)$ symmetric and $(642)$ asymmetric directions: Left panel shows maps of 10.2 nm, 37.0 nm and 92.5 nm Bi:YIG films from set-A. Right Panel shows maps of 100 nm and 200nm Bi:YIG films from set-B.}\label{F3-A}
\end{center}
\end{figure*}
Epitaxial Bi:YIG films were grown on GGG(111) crystals using a KrF Excimer laser (Lambda Physik COMPex Pro, $\lambda$ = 248 \si{\nano\meter}) of 20 \si{\nano\second} pulse width. The laser was fired at a repetition frequency of 10 \si{\Hz} on solid state synthesized $Bi_{0.25}Y_{2.75}Fe_{5}O_{12}$ target, placed 50 \si{\milli\meter} away from the substrate. The substrates were in-situ annealed at 800 $^\circ$C for 120 minutes to get atomically flat surfaces and then cooled down to 500 $^\circ$C in $4.0\times10^{-2}$  \si{\milli\bar} oxygen pressure to deposit the films. The target was sufficiently preablated before actual deposition to get a steady state target surface. We incorporated two routes to deposit these epitaxial films to obtain different strain-states by changing the laser fluence at a fixed oxygen ambient and growth temperature. For set-A, the fluence was $\sim$ 1 \si{\joule\per\centi\meter\squared}  with a spot size of $\sim$ 10.0 \si{\milli\meter\squared} and hence the realized growth rate was $\sim$ 0.25 \si{\AA\per\second}. For set-B, we almost doubled the fluence ($\sim$ 1.9 \si{\joule\per\centi\meter\squared}) by reducing the spot size ($\sim$ 5.4 \si{\milli\meter\squared}) to achieve an enhanced growth rate of $\sim$ 0.45 \si{\AA\per\second}. We deposited five films of thicknesses 10.2, 18.1, 37.0, 92.5 and 200 \si{\nano\meter} using set-A growth parameters, hereafter denoted as $A_{1}$, $A_{2}$, $A_{3}$, $A_{4}$ and $A_{5}$, respectively. Another five films of thicknesses 18.7, 39.8, 100, 150 and 200 \si{\nano\meter} were grown using growth protocol-B, hereafter denoted as $B_{1}$, $B_{2}$, $B_{3}$, $B_{4}$, and $B_{5}$ respectively. The growth rate and hence the thicknesses of different samples were pre-calibrated using Dektak stylus profilometer.
     PANalytical X’Pert PRO four circle diffractometer equipped with Cu-K$\alpha_{1}$ source ($\lambda = 1.54059$ \si{\AA}) was used to characterize the crystallinity and to quantify the state-of-strain. Room temperature Vibrating Sample Magnetometry (VSM) measurement was performed using a Quantum Design Physical Property Measurement System (PPMS). For the dynamic magnetization measurements, we used both commercial and a custom-made FMR setup. Angular dependent FMR measurements were performed using Bruker EMX EPR spectrometer with cavity mode frequency f $\cong$ 9.60 \si{\giga\Hz}. Frequency dependent FMR measurements were performed by using a broadband coplanar waveguide (CPW). The CPW assembly was housed in an external homogeneous DC magnetic field along with the superposition of a small and low frequency AC field. This small modulation of magnetic field is required to get differential of absorbed radio frequency (RF) power which is measured by a Schottky diode detector and a lock-in amplifier.\\
\begin{figure*}[t]
\begin{center}
\includegraphics[trim = 0mm 5mm 0mm 5mm, width=172mm]{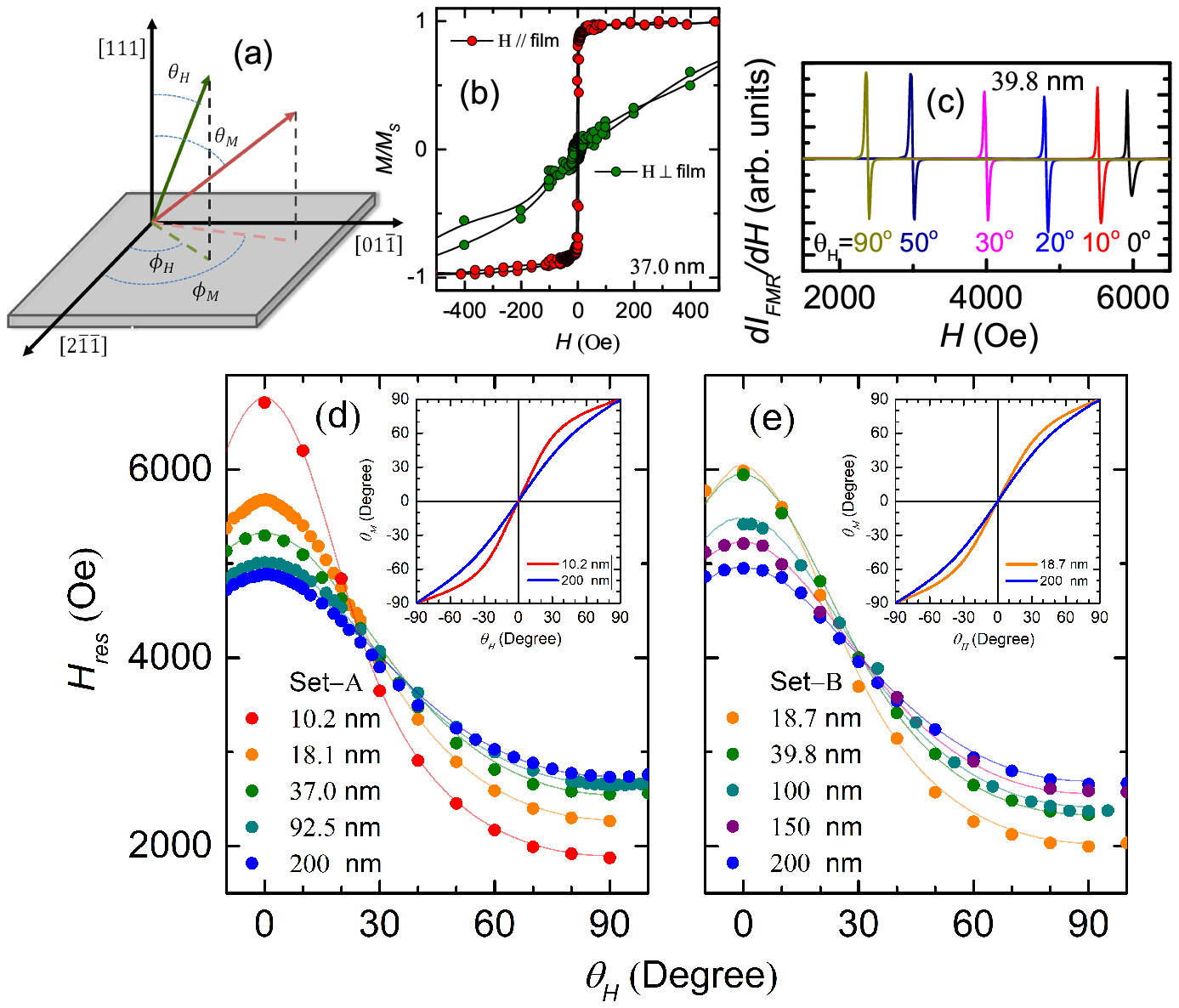}\
\setlength{\belowcaptionskip}{-5mm}
\caption{(Color online) Room temperature magnetization and out-of-plane angular dependence of resonance field for Bi:YIG films from both the sets. (a) Typical schematic of spherical coordinate system for FMR measurements and analysis of [111] oriented epitaxial Bi:YIG/GGG(111) samples. (b) Magnetic hysteresis loops measured in in-Plane (red) and out-of-plane (Green) configuration of a 37.0 \si{\nano\meter} thin film from set-A using VSM. (c) Representative FMR derivative spectra for a 39.8 \si{\nano\meter} Bi:YIG film from set-B. Panel (d) picturizes out-of-plane angular variation ($\theta_{H}$) of the resonance fields ($H_{res}$) and fitted curves for set-A. Inset: Energy minimization comparison for 10.2 \si{\nano\meter} and 200 \si{\nano\meter} thick Bi:YIG films from set-A. Panel (e) picturizes out-of-plane angular variation ($\theta_{H}$) of the resonance fields ($H_{res}$) and fitted curves for set-B. Inset: Energy minimization comparison  of 18.7 \si{\nano\meter} and 200 \si{\nano\meter} thick Bi:YIG films from set-B.}\label{F3}
\end{center}
\end{figure*}

 Fig. \ref{F2} summarizes the X-ray measurements on Bi:YIG films grown on (111) oriented GGG substrates. Panel (a) shows reflectivity measurements on all the samples except 100, 150 and 200 \si{\nano\meter} (pre-calibrated using profilometry), as there were no visible thickness fringes due to larger thickness. Reflectivity data was fitted to calculate and standardize the profilometric pre-calibrated thickness and gives very low roughness ranging from 0.25 to 0.39 \si{\nano\meter}. The panel (b) of Fig. \ref{F2} shows intensity normalized $\omega$ scan profiles with low values of full width half maximum (FWHM) ranging between 0.0448 to 0.0072 $^\circ$, signifies high crystallinity. The panel (d) of Fig. \ref{F2} shows X-ray diffraction patterns of all the Bi:YIG samples where the pronounced trail of Laue oscillations characterizes smooth surfaces and sharp interfaces. The bulk lattice constant for $Bi_{0.25}Y_{2.75}Fe_{5}O_{12}$ comes out to be 12.389 \si{\AA} and the corresponding 2$\theta$ peak position is shown by a vertical bar beneath substrate peak. Thin film lattice constant ($a_{\bot}$) differs due to lattice mismatch between substrate and film (shown by vertical up arrows). This lattice mismatch causes rhombohedral distortion in the films and hence contributes to diagonally stretched unit cells along the [111] growth direction. The strain induced rhombohedral distortion in these epitaxial Bi:YIG films can be quantified using the parameter $\sigma=(a_{b}-a_{\bot})/a_{b}=\Delta a/a_{b}$ , where, $a_{b}$ is the bulk Bi:YIG lattice parameter and $a_{\bot}$ is the stretched film lattice parameter along the [111] direction\cite{Wang2014}\cite{Du}\cite{Bhoi2018}. For set-A samples, XRD patterns show strain relaxation as the thickness increases from 10.2 to 200 \si{\nano\meter} (2$\theta$ value approaches the bulk value), the strain-induced lattice distortion decreases from $1.162\%$ to almost $\sim0.0\%$. Surprisingly, set-B samples having thicknesses 18.7, 39.8, 100, 150 and 200 \si{\nano\meter}, show relatively high strain ($1.122\%$ for 18.7 \si{\nano\meter} thin film and $0.171\%$ for 200 \si{\nano\meter} thick film). The variation of $a_{\bot}$ and the lattice strain ($\sigma$) w.r.t. to Bi:YIG film thickness from both the sets are shown in Fig. \ref{F2} panel (c). It can be seen that the value of $a_{\bot}$ approaches bulk value for a film of thickness 200 \si{\nano\meter} from set-A, whereas, a 200 \si{\nano\meter} thick film from set-B possess elongated $a_{\bot}$. Similarly, a 200 \si{\nano\meter} thick film from set-A show negligible lattice strain but a 200 \si{\nano\meter} thick film from set-B possesses reasonably large lattice strain. The 2-axis $\omega$ vs. $2\theta-\omega$ maps are shown in Fig. \ref{F3-A}. The top panel shows symmetric maps in the $444$ direction of Bi:YIG films. Whereas, the bottom panel shows the $642$ asymmetric direction maps. We show $(444)$ symmetric and $(642)$ asymmetric 2-axis maps for $10.2$, $37.0$ and $92.5$ \si{\nano\meter} films from set-A, and, $100$ and $200$ \si{\nano\meter} films from set-B. It can be clearly seen that the $2\theta-\omega$ value for film (represented by +; red colored) shifts toward higher value as the film thickness increases and approaches to the GGG substrate spot (represented by $\times$ ; black colored). The map of a $92.5$ \si{\nano\meter} thick film from set-A shows large relaxation compare to a $100$ \si{\nano\meter} thick film from set-B. Which confirms the inference drawn from $\theta-2\theta$ XRD measurement.              
 The laser ablation conditions greatly impact the lattice constant of deposited films irrespective of oxygen pressure and growth temperature. We observe that the laser fluence plays an important role in tuning the lattice constant of the films. The set-A films prepared using slow growth rate ($\sim$ 0.25 \si{\AA\per\second}) with a lower laser fluence ($\sim$ 1 \si{\joule\per\centi\meter\squared}) show less lattice expansion and complete relaxation with thickness increment. Whereas, the set-B films prepared using almost doubled growth rate ($\sim$ 0.45 \si{\AA\per\second}) due to higher laser fluence ($\sim$ 1.9 \si{\joule\per\centi\meter\squared}) show tendency to possess reasonably large lattice expansion even for higher thicknesses (panel (c) and (d) of Fig. \ref{F2}). The laser fluence (growth rate) is low in the case of Set-A Bi:YIG films, which gives sufficient settle down time to the ablated plasma species and hence lead to strain relaxation. In contrast, the higher laser fluence (growth rate) in the case of Bi:YIG films from set-B, doesn't allow the ablated plasma species to settle down and get relaxed.
 Table \ref{Table1} contains XRD, magnetization and FMR derived parameters for both the sets of samples. The negative sign of $\sigma$ indicates the presence of compressive strain which relaxes with increment in film thickness\cite{Wang2014,Bhoi2018,Kumar,Manuilov}.

\begin{table*}[htbp]
\caption{\label{tab:table1}XRD, $\textit{\textbf{M}}-\textit{\textbf{H}}$ and FMR derived parameters of Bi:YIG epitaxial films grown by two protocols. Set-A ($A_{1}$, $A_{2}$, $A_{3}$, $A_{4}$, and $A_{5}$) and set-B ($B_{1}$, $B_{2}$, $B_{3}$, $B_{4}$, and $B_{5}$)  are separated by a solid horizontal line.}
\begin{ruledtabular}
\begingroup
\renewcommand{\arraystretch}{1.2}
\begin{tabular}{ccccccccccc}
Thickness & $2\theta$ & $a_{\perp}$ & $\Delta a/a_{b}$ & $4\pi M_{S}$ & g-factor & $K_{u}$ & $H_{u}$ & $H_{1}$ & $H_{2}$ & $E_{ani}$\\

 &  &  &  & (VSM) &  &  &  &  &  & \\

(\si{\nano\meter}) & (Degree) & \si{\AA} & $\%$ & (Oe) &  & ($\times10^{3}$ erg/cc) & (Oe) & (Oe) & (Oe) & ($\times10^{3}$ erg/cc)\\
\hline\\
10.2 ($A_{1}$) & 50.406 & 12.533 & -1.162 & 1720$\pm$100 & 2.12 & -125.40$\pm$8.23 & -1831$\pm$227 & -15.3$\pm$1.9 & 3.1$\pm$1.1 & -126.24$\pm$8.24\\

18.1 ($A_{2}$) & 50.766 & 12.450 & -0.492 & 1432$\pm$63 & 2.09 & -55.68$\pm$4.21 & -977$\pm$117 & -29.7$\pm$2.2 & 13.8$\pm$1.6 & -56.59$\pm$4.21 \\

37.0 ($A_{3}$) & 50.919	& 12.41 & -0.210 & 1482$\pm$37 & 2.03 & -27.68$\pm$3.02 & -469$\pm$63 & -40.9$\pm$1.8 & 30.9$\pm$1.7 & -28.27$\pm$3.01\\

92.5 $(A_{4}$) & 51.011 & 12.394 & -0.0484 & 1507$\pm$38 & 2.01 & -7.43$\pm$2.71 & -124$\pm$48 & -12.1$\pm$0.9 & 52.1$\pm$2.0 & -5.03$\pm$2.70\\

200 ($A_{5}$) & 51.033 & 12.389 & 0.0 & 1407$\pm$25 & 2.00 & -3.91$\pm$1.32 & -70$\pm$25 & -3.2$\pm$0.6 & 3.2$\pm$0.7 & -3.91$\pm$1.31\\
\\
\hline\\
18.7  ($B_{1}$) & 50.425 & 12.528 & -1.122 & 1582$\pm$38 & 2.13 & -81.42$\pm$6.72 & -1292$\pm$137 & -57.3$\pm$1.9 & 6.7$\pm$1.0 & -84.61$\pm$6.81\\

39.8  ($B_{2}$) & 50.530 & 12.504 & -0.928 & 1545$\pm$25 & 2.05 & -62.12$\pm$5.33 & -1010$\pm$103 & -14.4$\pm$0.8 & 160.2$\pm$3.4 & -53.16$\pm$5.42\\

100 ($B_{3}$) & 50.747 & 12.454 & -0.525 & 1520$\pm$25 & 2.04 & -39.23$\pm$4.28 & -648$\pm$82 & -40.0$\pm$1.2 & 65.1$\pm$1.9 & -37.71$\pm$4.37\\

150 ($B_4$)	& 50.807 & 12.440 & -0.414 & 1608$\pm$17 & 2.03 & -15.78$\pm$3.04 & -246$\pm$50 & -19.6$\pm$0.6 & 118.5$\pm$1.4 & -9.45$\pm$3.11\\

200 ($B_{5}$) & 50.939 & 12.410 & -0.171 & 1457$\pm$12 & 2.01 & -6.25$\pm$0.91 & -108$\pm$17 & -23.9$\pm$0.7 & 113.5$\pm$1.2 & -1.06$\pm$0.96\\
\end{tabular}
\endgroup
\end{ruledtabular}
\label{Table1}
\end{table*}

Room temperature in-plane magnetic hysteresis loops are measured using VSM on Quantum Design PPMS. In-plane and out-of-plane magnetization loops for a 37.0 \si{\nano\meter} thick film from set-A is shown in Fig. \ref{F3}(b), where the paramagnetic background from GGG was subtracted. The values of saturation magnetization ($4\pi M_{S}$) for samples from set-A and set-B ranges between 1720$\pm$100 to 1407$\pm$25 Oe and 1608$\pm$17 to 1457$\pm$12 Oe, respectively. The coercivity ($H_{C}$) of these samples are in the range of $\sim$ 13 to 23 Oe. These values fall in the range of reported YIG magnetization data\cite{Onbasli2014}\cite{Bhoi2018,Kumar,Manuilov,Dorsey,Ibrahim}. To probe the static and dynamic magnetic properties of Bi:YIG epitaxial films, we performed angular and frequency dependent FMR measurements on both the sets of samples. Generally, the magnetic garnet thin films with a hard axis in the [111] direction (i.e., In-plane easy axis), possesses extrinsic uniaxial magnetic and intrinsic magnetocrystalline cubic anisotropies. FMR can directly deduce the magnetic anisotropies in a precise manner. The coordinate system used for FMR study on (111) oriented epitaxial Bi:YIG films is shown in Fig. \ref{F3}(a). The orientations of static magnetic field $\textit{\textbf{H}}$ and magnetization vector $\textit{\textbf{M}}$ with reference to coordinates x:[2$\overline{1}$ $\overline{1}$], y:[01$\overline{1}$] and z:[111] are described by the angles $\phi_{H}$, $\theta_{H}$ and $\phi_{M}$, $\theta_{M}$, respectively. The total free energy per unit volume of the media for (111) oriented cubic garnet system has the form\cite{Landau}\cite{Lee2016_JAP},
	
\begin{widetext}	
\begin{equation}\label{eq1}
\begin{array}{c}
F =  - H{M_S}\left[ \begin{array}{l}
\sin {\theta _H}\sin {\theta _M}cos\left( {{\phi _H} - {\phi _M}} \right)\\
 + \cos {\theta _H}\cos {\theta _M}
\end{array} \right] + 2\pi M_S^2{\cos ^2}{\theta _M} - {K_u}{\cos ^2}{\theta _M} + \frac{{{K_1}}}{{12}}\left( \begin{array}{l}
7{\sin ^4}{\theta _M} - 8{\sin ^2}{\theta _M} + 4 - \\
4\sqrt 2 {\sin ^3}{\theta _M}\cos {\theta _M}\cos 3{\phi _M}
\end{array} \right)\\
 + \frac{{{K_2}}}{{108}}\left( { - 24{{\sin }^6}{\theta _M} + 45{{\sin }^4}{\theta _M} - 24{{\sin }^2}{\theta _M} + 4 - 2\sqrt 2 {{\sin }^3}{\theta _M}\cos {\theta _M}\left( {5{{\sin }^2}{\theta _M} - 2} \right)\cos 3{\phi _M} + {{\sin }^6}{\theta _M}\cos 6{\phi _M}} \right)
\end{array}\
\end{equation}
\end{widetext}

\begin{figure}[t]
\begin{center}
\includegraphics[trim = 5mm 3mm 0mm 2mm, width=86mm]{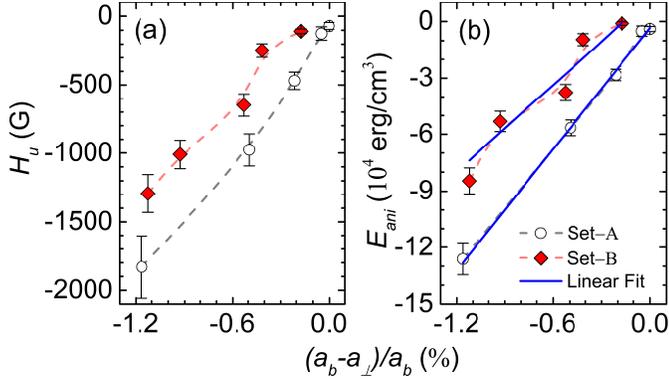}\
\setlength{\belowcaptionskip}{-5mm}
\caption{(Color online) (a) Out-of-plane uniaxial anisotropy field $H_{u}$ and (b) total anisotropy energy $E_{ani}$ as a function of the rhombohedral distortion (($a_{b}-a_{\bot}))/a_{b}\%$  of the Bi:YIG films on GGG(111). Blue solid lines are the least-square fit to obtain magnetoelastic coupling constant. Dashed curves serve as a guide to the eye.}\label{F4}
\end{center}
\end{figure}

The first term in Eq. \eqref{eq1} corresponds to the Zeeman energy, the second term to the demagnetization energy, the third term to the out-of-plane uniaxial magnetocrystalline anisotropy energy $K_{u}$ and the last two terms are due to first and second order cubic magnetocrystalline anisotropy energies, $K_{1}$  and $K_{2}$, respectively. The total free energy equation was minimized ($\partial{F}/\partial{\theta_{M}}\equiv\partial{F}/\partial{\phi_{M}}\equiv0$) to obtain the equilibrium orientation of the magnetization vector $\textit{\textbf{M}}(\textit{\textbf{H}})$. The evaluation of resonance frequency ($\omega_{res}$) of uniform magnetization precessional mode at equilibrium condition can be made using total free energy and is expressed as:\cite{Lee2016_JAP,Suhl,Smit}

\begin{equation}\label{eq2}
{\omega _{res}} = \frac{\gamma }{{{M_S}\sin {\theta _M}}}{\left[ {\frac{{{\partial ^2}F}}{{\partial \theta _M^2}}\frac{{{\partial ^2}F}}{{\partial \phi _M^2}} - {{\left( {\frac{{{\partial ^2}F}}{{\partial {\theta _M}\partial {\phi _M}}}} \right)}^2}} \right]^{{\raise0.7ex\hbox{$1$} \!\mathord{\left/
 {\vphantom {1 2}}\right.\kern-\nulldelimiterspace}
\!\lower0.7ex\hbox{$2$}}}}\
\end{equation}

here $\gamma$ and $M_{S}$ denote gyromagnetic ratio and saturation magnetization, respectively. These coupled and indirectly defined functional equations were solved numerically to obtain the equilibrium angles at resonance condition and fit the angular dependent resonance data ($H_{res}$ vs. $\theta_{H}$) to determine g-factor, $K_{u}$, $H_{u}$, $H_{1}$, $H_{2}$ and $E_{ani}$ (see Table I). Fig. \ref{F3} (c) shows representative angular-FMR spectra of a 39.8 \si{\nano\meter} thick film from set-B at a microwave frequency of $\sim$ 9.6 \si{\giga\Hz}. The peak-to-peak difference of FMR derivative gives linewidth ($\Delta H$) which decreases as the film thickness increases. The measured in-plane $\Delta H$ values for set-A samples $A_{1}$, $A_{2}$, $A_{3}$, $A_{4}$, and $A_{5}$ at $\sim$ 9.6 \si{\giga\Hz} are 154, 120, 93, 39, and 14 Oe, respectively. Similarly, for set-B samples  $B_{1}$, $B_{2}$, $B_{3}$, $B_{4}$, and $B_{5}$ the in-plane $\Delta H$ values are 150, 105, 50, 44, and 23 Oe, respectively. The energy minimization governed by the correspondence between  $\theta_{H}$ and  $\theta_{M}$ is shown in the inset of Fig. \ref{F3}(d-e), where the equilibrium magnetization angle  $\theta_{M}$ was estimated numerically.  It can be seen that energy minimization attains large curvature for thin Bi:YIG film from both the sets and hence large anisotropy compare to thick film from the respective sets. Fig. \ref{F3}(d) and (e) show $\theta_{H}$ dependence of $H_{res}$ for set-A and set-B samples, respectively. The fit using Eqs. \eqref{eq1} and \eqref{eq2} agrees well with the measured data. All the extracted parameters for both the sets of samples are shown in Table \ref{Table1}, separated by a solid line.\\
 We mainly focus on the out-of-plane uniaxial anisotropy field ($H_{u}$) due to its large contribution to total magnetic anisotropy and systematic variation with film thickness or lattice strain. In contrast, we couldn’t witness a systematic thickness or strain dependence of cubic first and second order anisotropy which are weak in magnitude. Interestingly, $H_{u}$ for 10.2 \si{\nano\meter} thin and 200 \si{\nano\meter} thick films from set-A comes out to be -1831$\pm$227 Oe and -70$\pm$25 Oe, respectively, which provides a strain tuning over a range of more than 1700 Oe. It suggests that the rhombohedral distortion induces substantial out-of-plane uniaxial anisotropy via the magnetostriction, which decreases systematically with increase in the film thickness. The g-factor for thin films is as large as 2.13, greater than the spin-only value $2.0$. This corroborates the existence of spin-orbit coupling that lead to strain-induced anisotropy. However, the g-factor for thick films are smaller ($\sim 2.0$). This variation in ‘g’ possibly arises due to different strain state, which may change the occupation of orbitals and hence the magnitude of orbital angular momentum and spin-orbit coupling. The strain induced variation of $H_{u}$ and $E_{ani}$ is picturized in Fig. \ref{F4} (a) and (b), respectively. It is clear from Fig. \ref{F4} (a) and (b) that the magnitudes of $H_{u}$ and $E_{ani}$ increases almost linearly as the magnitude of rhombohedral distortion increases. The enhancement in uniaxial anisotropy field is due to the larger magnitude of growth induced strain in the samples from set-B as compare to set-A. The substrate-film lattice mismatch causes lattice-distortion in deposited films which results in a definite strain-state. The lattice distortion influences the magnetic properties. This magnetization-lattice coupling gives rise to strain-induced out-of-plane uniaxial anisotropy field, $H_{u}$. The strain induced by rhombohedral distortion in a cubic lattice relaxes as the film thickness increases and hence results in very low or almost negligible strain, which ultimately makes the film isotropic, having properties similar to bulk. The value of $H_{u}$ for Bi:YIG films $B_{1}$, $B_{2}$, $B_{3}$, $B_{4}$, and $B_{5}$ from set-B are found to be -1292$\pm$137, -1010$\pm$103, -648$\pm$82, -246$\pm$50, and -108$\pm$17 Oe, respectively. It is important to note that the values of $H_{u}$ for thicker films from set-B are larger compare to respective film thicknesses from set-A. If we compare the uniaxial anisotropy field of Bi:YIG films from both the sets of almost equal thicknesses, i.e., $A_{2}$ (18.1 \si{\nano\meter}) and $B_{1}$ (18.7 \si{\nano\meter}), comes out to be -977$\pm$117 Oe, and -1292$\pm$137 Oe, respectively. The uniaxial anisotropy field magnitude for set-B Bi:YIG film is almost 300 Oe larger compare to the value of set-A Bi:YIG film. 

 The magnetoelastic energy density for a strain dependent FMR measurement is given by $F_{ME} = -\sigma b[cos]^{2}\Theta$, where $b$ is magnetoelastic constant and $\Theta$ is the angle between $\textit{\textbf{M}}$ and strain direction\cite{Wang2014}\cite{Du}. For $\textit{\textbf{M}}$ pointing in the [111] direction, the magnetoelastic energy density has the form, $F_{ME} = -\sigma b$. Fig. \ref{F4}(b) shows the linear dependence and least-square fit of anisotropy energy $E_{ani}= - 1/2[M_{S}H_{u}]$ with different strain states of Bi:YIG films from both the sets. The derived expressions from least-square fit in Fig. \ref{F4}(b) for set-A and set-B are $E_{ani}=(-3.46\pm1.06)\times10^{3}+(10.74\pm0.51)\times10^{6}[(a_{b}-a_{\perp})/a_{b}]$ (erg/cc) and $E_{ani}=(12.58\pm3.59)\times10^{3}+(7.71\pm1.18)\times10^{6}[(a_{b}-a_{\perp})/a_{b}]$ (erg/cc), where the slope of the lines give $-b=(10.74\pm0.51)\times10^{6}$ (erg/cc) and $-b=(7.71\pm1.18)\times10^{6}$ (erg/cc), respectively. The negative sign of $b$ implies that the magnetic easy axis is parallel to the compressed lattice plane; [111]. The magnetoelastic constant of Bi:YIG comes out to be larger than in pure-YIG film\cite{Wang2014}. Pure YIG exhibits almost quenched orbital momentum of half-filled \textit{d} shell in $Fe^{3+}$ electron configuration, leads to weak SOC and shows low magnetoelastic coupling constant. The substitution of strong SOC ions such as $Bi^{3+}$, $Dy^{3+}$ and $Tm^{3+}$ etc. enhances the spin-orbit coupling which results in improved magnetoelastic coupling. It suggests that the strain-tuning could be very crucial to obtain large magnetocrystalline anisotropy even in thick ferrimagnetic-insulating films. 
 
 \begin{figure}[t]
\begin{center}
\includegraphics[trim = 0mm 5mm 0mm 5mm, width=86mm]{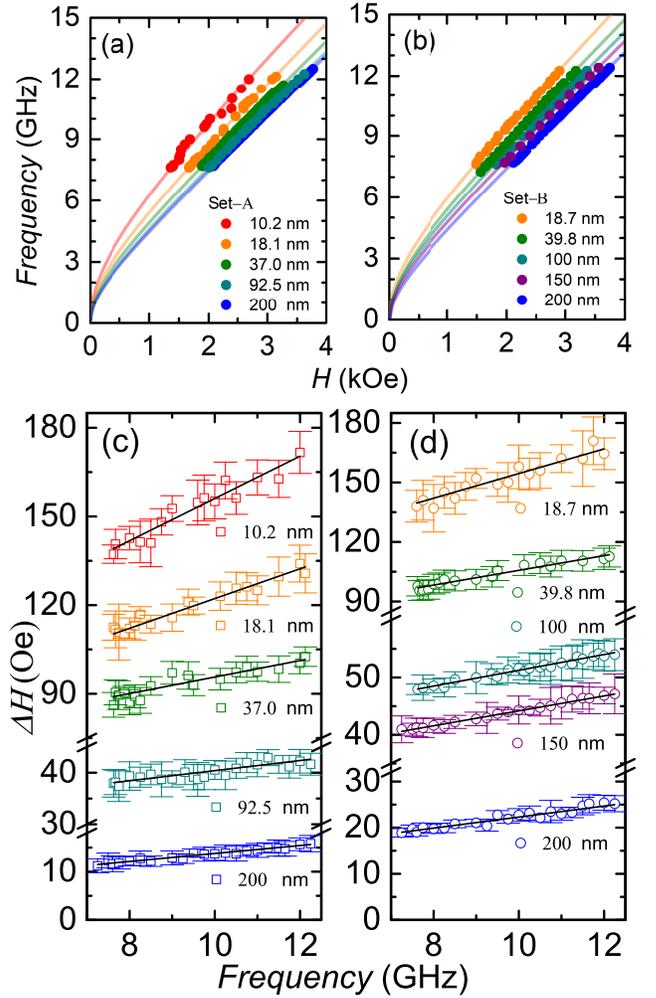}\
\setlength{\belowcaptionskip}{-7mm}
\caption{(Color online) In-plane, frequency and thickness dependent room temperature FMR measurements. Panel (a) and (b) represent frequency vs. resonance field plots for set-A and set-B, respectively. The fit to experimental data has been shown by corresponding coloured solid curves. Panel (c) and (d) represent frequency dependent linewidth variation for set-A and set-B, respectively. Black solid lines represent fit to the experimental data.}\label{F5}
\end{center}
\end{figure}	
	
	Gilbert damping coefficient $\alpha$ for our Bi:YIG films has been calculated from frequency-dependent FMR measurement between 7 and 12 \si{\giga\Hz}. The external magnetic field is swept at various fixed frequencies. Fig. \ref{F5} (a) and (b) show the frequency vs. $H_{res}$ data and its fit (corresponding colored solid curves) for set-A and set-B, respectively, using reduced form of Eqs. \eqref{eq1} and \eqref{eq2} in a limiting in-plane magnetic field geometry ($\theta_{H}=90^{\circ}$,$\phi_{H} = 0^{\circ}$). The derived compact expression in asymptotic limit has the form (in-plane Kittel equation),
\begin{equation}\label{eq3}
{\omega _{res}} = \gamma \sqrt {{H_{res}}\left( {{H_{res}} + 4\pi {M_{eff}}} \right)} \
\end{equation}
with the effective magnetization $4\pi M_{eff}= 4\pi M_{S}-H_{ani} $, where, $H_{ani}$ is the anisotropy field parameterizes out-of-plane uniaxial and cubic anisotropies. It is clear from Fig. \ref{F5} (a) and (b) that the data fits perfectly without even considering additional in-plane anisotropy contributions. In Eq. \eqref{eq3} we do not consider a renormalization shift in the resonance frequency and a small shift in resonance field which can arise by two-magnon scattering and a static dipole interaction between the ferrimagnetic film and the paramagnetic substrate, respectively, due to negligibly small contributions. Fig. \ref{F5} (c) and (d) show in-plane frequency dependencies of linewidth ($\Delta H$) for set-A and set-B films, respectively. The standard Landau-Lifshitz-Gilbert equation justifies the linear dependence of $\Delta H$ with frequency and used for straightforward determination of the intrinsic Gilbert damping coefficient ($\alpha$): $\Delta H=\Delta H_{0}+(4\pi\alpha/\sqrt{3}\gamma)f_{res}$, where $\Delta H_{0}$ is the extrinsic linewidth broadening due to magnetic inhomogeneities within the material. The extracted values of $4\pi M_{eff}$, $\alpha$ and $\Delta H_{0}$ for films from both the sets are shown in table \ref{Table2}.

\begin{table}[htbp]
\setlength{\abovecaptionskip}{0mm}
\caption{\label{tab:table2}Frequency and thickness dependent FMR derived effective magnetization, Gilbert damping coefficient and inhomogeneous broadening of Bi:YIG epitaxial films grown by two different protocols. Set-A and set-B are separated by a solid horizontal line.}
\begin{ruledtabular}
\begingroup
\setlength{\tabcolsep}{6pt} 
\renewcommand{\arraystretch}{1.2}
\begin{tabular}{ccccc}
Thickness & $4\pi\ M_{eff}$  & $\alpha (\times10^{-3})$ & $\Delta H_{0}$\\
   
(\si{\nano\meter}) & (Oe)&   &(Oe) \\
\hline
10.2 ($A_{1}$)& 3482$\pm$65   & 18.3$\pm$1.3 & 84\\
18.1 ($A_{2}$) & 2441$\pm$27  & 12.7$\pm$0.9 & 72\\
37.0 ($A_{3}$) & 1970$\pm$8   & 6.9$\pm$0.7 & 67\\
92.5 ($A_{4}$) & 1673$\pm$46  & 2.4$\pm$0.3 & 30\\
200 ($A_{5}$) & 1510$\pm$2    & 2.0$\pm$0.1 & 5\\
\hline
18.7  ($B_{1}$) & 2928$\pm$15  & 16.1$\pm$1.5 & 92\\
39.8 ($B_{2}$) & 2437$\pm$2    & 9.6$\pm$0.6 & 68\\
100 ($B_{3}$) & 2125$\pm$3     & 3.4$\pm$0.1 & 37\\
150 ($B_{4}$) & 1787$\pm$4     & 3.2$\pm$0.1 & 31\\
200 ($B_{5}$) & 1399$\pm$2     & 2.9$\pm$0.2 & 10\\
\end{tabular}
\endgroup
\end{ruledtabular}
\label{Table2}
\end{table}

\begin{figure}[t]
\begin{center}
\includegraphics[trim = 0mm 3mm 0mm 1mm, width=80mm]{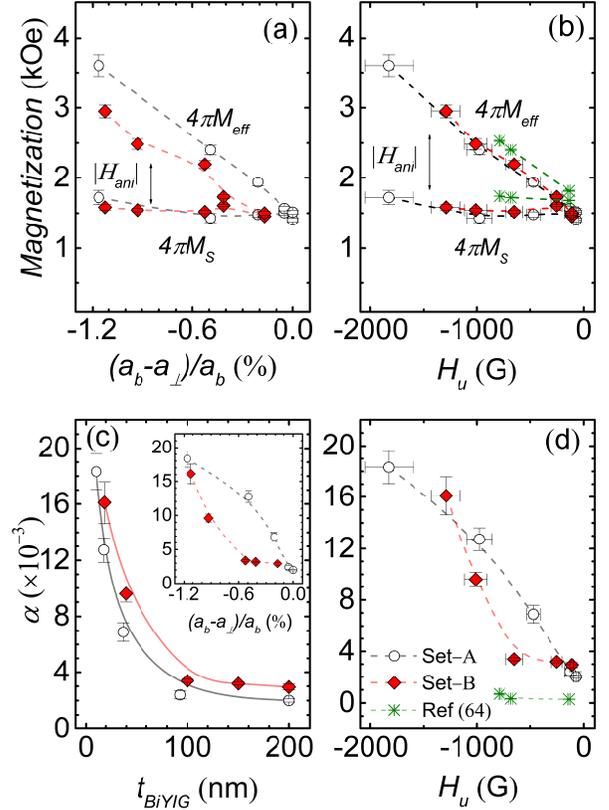}\
\setlength{\belowcaptionskip}{-7mm}
\caption{(Color online) Magnetization ($4\pi M_{S}$ and $4\pi M_{eff}$) dependencies on (a) epitaxial strain and (b) $H_{u}$. Precessional damping dependencies on (c) thickness (inset: on strain) and (d) $H_{u}$. Panel (b) and (d) include YIG/GGG(111) data from ref.\cite{Bhoi2018}. Dashed curves serve as a guide to the eye.}\label{F6}
\end{center}
\end{figure}

Fig. \ref{F6} (a) shows strain dependent variations of $4\pi M_{eff}$ and $4\pi M_{S}$. The values of $4\pi M_{eff}$ for both the sets systematically decreases with the increase in film thickness but the values strongly depend on the state-of-the-strain in the films. It can be seen that $4\pi M_{eff}$ is significantly larger than the Bi:YIG saturation magnetization generated simple shape anisotropy i.e., $4\pi M_{S}$, revealing the presence of a negative uniaxial anisotropy, signature of easy in-plane magnetization. The gap between $4\pi M_{eff}$ and $4\pi M_{S}$ represents magnitude of anisotropy field $H_{ani}=4\pi M_{S}-4\pi M_{eff}$ which decreases with increment in film thickness. The magnitude of $H_{ani}$ for $\sim$ 100 \si{\nano\meter} thick Bi:YIG film from set-B is larger than that expected and comparable to $\sim$ 37 \si{\nano\meter} thin film from set-A, which is due to growth induced large strain. Fig. \ref{F6}(b) shows magnetization ($4\pi M_{eff}$, $4\pi M_{S}$) dependence on uniaxial anisotropy field, where, the magnetization decreases in proportion with the magnitude of uniaxial anisotropy field. Fig. \ref{F6} (c) shows the variation of $\alpha$ with respect to the film thickness from both the sets. Whereas, inset shows induced strain dependency of $\alpha$. We notice that the value of $\alpha$ decreases nonlinearly as film thickness increases (or strain relaxes) and vice-versa. We include effective magnetization, uniaxial anisotropy field and damping data of YIG/GGG(111) films from literature by Bhoi \textit{et. al.}\cite{Bhoi2018} which also follow the same trend. The lowest damping possessed by a 200 \si{\nano\meter} thick film from set-A is ($2.0\pm0.1)\times10^{-3}$ with an inhomogeneous broadening of $\sim$ 6 Oe, whereas, a 200 \si{\nano\meter} thick film from set-B shows slightly larger damping ($2.9\pm0.2)\times10^{-3}$ with an inhomogeneous broadening of $\sim$ 10 Oe but inherit reasonably large uniaxial anisotropy field ($-108\pm17$ Oe) which is almost two times larger compare to former. Although, the damping in Bi doped YIG enhances due to strong spin orbit coupling, still it’s passably small compare to metallic systems\cite{Okada2014,Song,Guo2014}. As the values of $\alpha$ and $\mid H_{u}\mid$ increases as a function of the induced strain, we therefore plot $\alpha$ vs. $H_{u}$ graph (see Fig. \ref{F6} (d)) to see the correlation between the precessional damping and magnetic anisotropy. In our Bi:YIG thin film system, we observe a nonlinear relationship between $\alpha$ and $H_{u}$, similar to YIG and can be attributed to spin wave damping induced by increment in strain\cite{Bhoi2018}. Rhombohedral distortion arising due to lattice mismatch between the film and the substrate leads to change in magnetic properties through spin orbit coupling\cite{Du}. The inclusion of lattice distorted SOC along with phonon-magnon scattering, two-magnon scattering or charge transfer relaxation may explain the thickness dependent enhancement of uniaxial anisotropy and reduction of magnetic damping\cite{Wang2014,Du,Jermain,Song,Guo2014,Bhoi2018,Liu2014}.

In summary, we have been able to grow high quality epitaxial Bi:YIG thin films on GGG(111) crystals as evidenced by prominent Laue oscillations in X-ray diffraction pattern. A usual trend of the film lattice relaxation and decrease in magnetic anisotropies as the film thickness increases has been observed. Our study shows that strain can be a crucial parameter to tune the magnetocrystalline anisotropy.  We optimize a growth protocol to get thick epitaxial films with large lattice strain which allows us to achieve large magneto-crystalline anisotropy. The Bi:YIG films grown using higher laser fluence show large magneto-crystalline anisotropy compare to films of respective thicknesses grown using lower laser fluence.  We show that the incorporation of growth induced large strain in thick Bi:YIG films can be helpful to improve the magnetic properties. Out-of-plane uniaxial anisotropy varies linearly with strain induced rhombohedral distortion of Bi:YIG lattice. Still, we are able to achieve fairly low Gilbert damping $\sim$ $2\times10^{-3}$ with enhanced magnetoelastic coupling. Further, as Bismuth substitution enhances the magneto-optical responses enormously, the coupling of large magnetocrystalline anisotropy, improved magnetoelastic coupling and low damping with strong magneto-optical activity in Bismuth substituted YIG may provide unique opportunities for photon-based-magnonics to develop efficient and low loss spintronics and caloritronics devices.\\

\textbf{ACKNOWLEDGEMENTS:}\\

              We thank Prof. R. C. Budhani for fruitful discussion and Dr. Veena Singh for technical assistance during FMR measurements. \\


\newpage
\newpage

\textbf{Reference details of the relationship between effective magnetization and Gilbert damping coefficient shown in Fig. 1.} It was constructed using the effective magnetization (saturation magnetization in few cases) and Gilbert damping coefficient values from various (Region I and II) ferro- and ferrimagnetic insulators, (Region III) conducting oxides and (Region IV) pure metals and metal-alloys, as reported in previous studies.\\

\begin{figure*}[t]
\begin{center}
\includegraphics[trim =0mm 5mm 0mm 5mm, width=172mm]{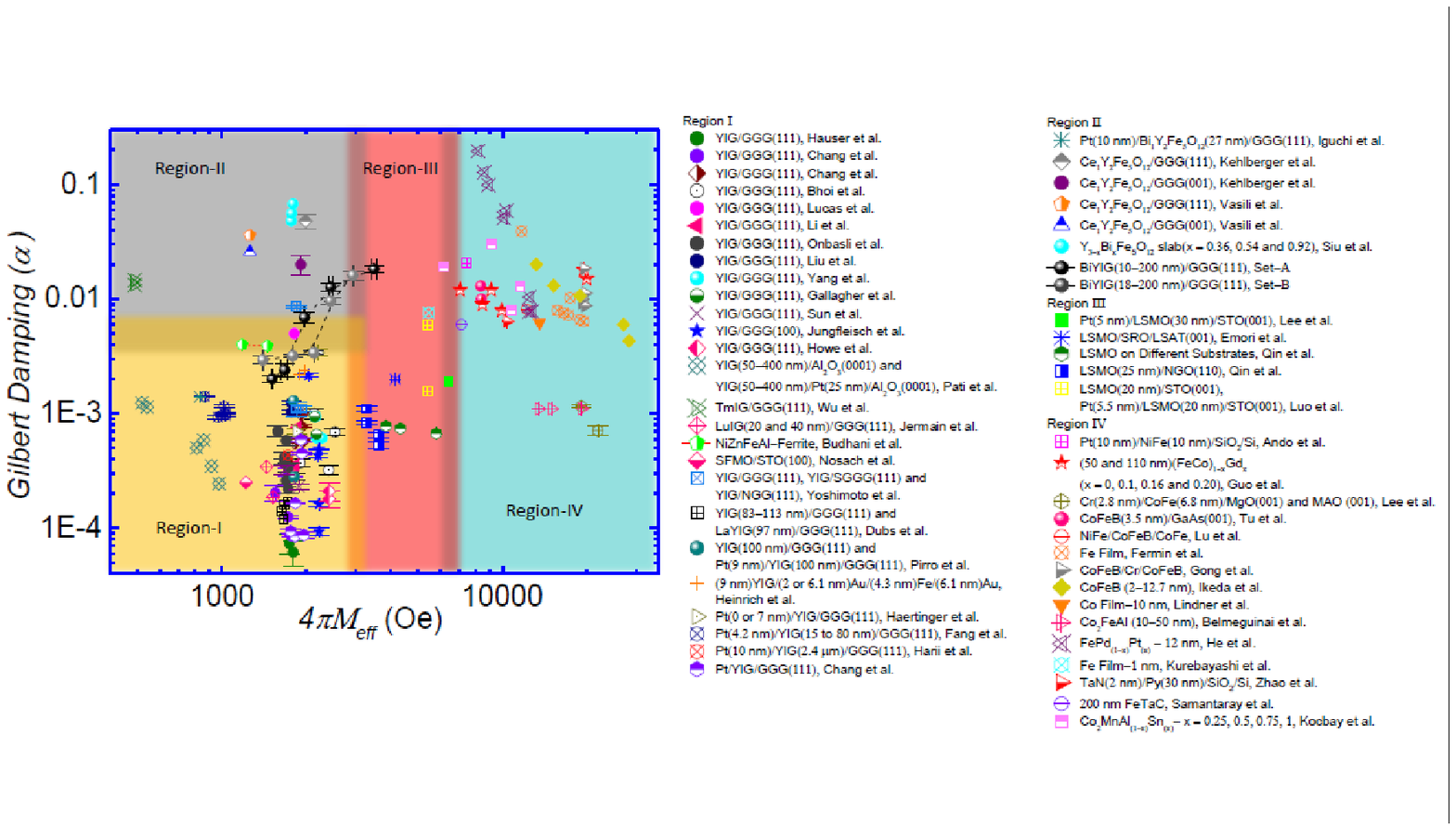}
\caption{Region - I: Hauser \textit{et al.}\cite{Hauser}, Chang \textit{et al.}\cite{Chang2014}, Chang \textit{et al.}\cite{Chang2017}, Bhoi \textit{et al.}\cite{Bhoi2018}, Lucas \textit{et al.}\cite{Lucas2017}, Le \textit{et al.}\cite{Li2016}, Onbasli \textit{et al.}\cite{Onbasli2014}, Liu \textit{et al.}\cite{Liu2014},  Yang\textit{et al.} \cite{Yang2018}, Gallagher \textit{et al.}\cite{Gallagher}, Sun \textit{et al.}\cite{Sun2012}, Jungfleish \textit{et al.}\cite{Jungfleisch}, Howe \textit{et al.}\cite{Howe}, Pati \textit{et al.}\cite{Pati}, Wu \textit{et al.}\cite{Wu2018}, Jermain \textit{et al.}\cite{Jermain}, Budhani \textit{et al.}\cite{Budhani2018}, Nosach \textit{et al.}\cite{Nosach}, Yoshimoto \textit{et al.}\cite{Yoshimoto}, Dubs \textit{et al.}\cite{Dubs}, Pirro \textit{et al.}\cite{Pirro}, Heinrich \textit{et al.}\cite{Heinrich}, Haertinger \textit{et al.}\cite{Haert2015}, Fang \textit{et al.}\cite{Fang2017}, Harii \textit{et al.}\cite{Harii}, Chang \textit{et al.}\cite{Chang2017}.\\
\\
Region - II: Iguchi \textit{et al.}\cite{Iguchi}, Kehlberger \textit{et al.}\cite{Kehlberger}, Vasili \textit{et al.}\cite{Vasili}, Siu \textit{et al.}\cite{Siu}.\\
\\
Region - III: Lee \textit{et al.}\cite{Lee2016_AIP}, Emori \textit{et al.}\cite{Emori2016}, Qin \textit{et al.}\cite{Qin2017}, Qin \textit{et al.}\cite{Qin2018}, Luo \textit{et al.}\cite{Luo2015}.\\
\\
Region - IV: Ando \textit{et al.}\cite{Ando}, Guo \textit{et al.}\cite{Guo2014}, Lee \textit{et al.}\cite{Lee2017}, Tu \textit{et al.}\cite{Tu2017}, Lu \textit{et al.}\cite{Lu2014}, Fermin \textit{et al.}\cite{Fermin}, Gong \textit{et al.}\cite{Gong}, Ikeda \textit{et al.}\cite{Ikeda2010}, Lindner \textit{et al.}\cite{Lindner}, Belmeguenai \textit{et al.}\cite{Belmeg2013}, He \textit{et al.}\cite{He}, Kurebayashi \textit{et al.}\cite{Kureb2013}, Zhao \textit{et al.}\cite{Zhao}, Samantaray \textit{et al}.\cite{Samantaray}, Kocbay \textit{et al.}\cite{Kocbay2012}.} 
\end{center}
\end{figure*}

\newpage
\newpage
\clearpage
\bibliographystyle{apsrev4-1}
\bibliography{Reference_BiYIG}
\end{document}